\documentclass[12pt]{article}
\usepackage[utf8]{inputenc}

\usepackage{newtxtext,newtxmath}

\usepackage{graphicx}

\usepackage[letterpaper,margin=1in]{geometry}

\renewenvironment{abstract}
	{\quotation}
	{\endquotation}

\date{}

\makeatletter
\renewcommand{\fnum@figure}{\textbf{Figure \thefigure}}
\renewcommand{\fnum@table}{\textbf{Table \thetable}}
\makeatother

\usepackage{scicite}

\usepackage{url}

\def\scititle{
	ND1 centers in diamond for long-term data storage in extreme conditions
}
\title{\bfseries \boldmath \scititle}

\author{
    Ahsan Ali$^1$,
    Wei Huang$^{1,2}$,
    Khadga Jung Karki $^{1,3\ast}$\and
	\small$^{1}$Department of Physics, Guangdong Technion-Israel Institute of Technology, Shantou \& Postal Code, China.\and
	\small$^{2}$Department of Chemistry, Techion-Israel Institute of Technlogy, Haifa \& Postal Code, Israel.\and
    \small$^{3}$ Matec, Guangdong Technion-Israel Institute of Technology, Shantou \& Postal Code, China.\and
	\small$^\ast$Corresponding author. Email: khadga.karki@gtiit.edu.cn
}

\begin{document} 

\maketitle

\begin{abstract} \bfseries \boldmath
Practically feasible long-term data storage under extreme conditions is an unsolved problem in modern data storage systems. This study introduces a novel approach using ND1 centers in diamonds for high-density, three-dimensional optical data storage. By employing near-infrared femtosecond laser pulses, we demonstrate the creation of sub-micron ND1 defect sites with precise spatial control, enabling efficient data encoding as luminescent ``pits." The ND1 centers exhibit robust photoluminescence in the UV spectrum, driven by three-photon absorption, which intrinsically provides a 3D reading of the data. Remarkably, these centers remain stable under extreme electric and magnetic fields, temperatures ranging from 4 K to 500 K, and corrosive chemical environments, with no degradation observed over extended periods. A reading speed of 500 MBits/s, limited by the lifetime of the photoluminescence, surpasses conventional Blu-ray technology while maintaining compatibility with existing optical data storage infrastructure. Our findings highlight diamond-based ND1 centers as a promising medium for durable, high-capacity data storage, capable of preserving critical information for millions of years, even under harsh conditions.

\end{abstract}

\section*{INTRODUCTION}\noindent
Digital transformation is experiencing a significant increase in data volume, with an exponential increase in zettabytes annually. The ever-growing amount of data has led to a proportional increase in the demand for storage devices~\cite{CAO2014}. Direct area storage devices, a term coined in 1964, encompass all formats, including optical, electronic, and magnetic ones. Optical storage options, such as compact discs (CDs), digital video discs (DVDs), and Blu-ray discs, were once considered the most reliable, with capacities of up to 300 GB and lifespans of nearly 30 years. However, the capacity margins of optical data storage (ODS) discs have pushed end users toward mediums such as flash drives and solid-state drives (SSD)~\cite{JANSEN2011}. In addition, thousands of data centers have been established to accommodate large amounts of data from commercial consumers to research institutions. Data centers rely on solid-state drives with a 10-year lifespan to protect their data, which means that they need to replace them regularly, incurring additional costs~\cite{FULLERTON2014}. In this context, long-life-span data storage systems are sought to archive high-value critical data by eliminating resource-intensive reading and writing during periodic changes in storage media. 

The development of next-generation ODS devices focuses on key parameters, such as capacity, speed, and extended life. Various media, such as plasmonic metal nanoparticles, semiconductor quantum dots, holographic memory, photochromic materials, and polymer composites, have been used to increase the capacity of two-dimensional storage systems. At the same time, the feasibility of multidimensional optical data storage (ODS) has been reported~\cite{LAMON2016,GU2009,JIN2014,XIANGPING2021,WANG2023}, for which color centers in diamonds are a promising option. Color centers, mainly single nitrogen vacancies (NV), are commonly generated in diamonds by ion implantation~\cite{WRACHTRUP2005}, electron irradiation~\cite{KOHLER1999}, and chemical vapor deposition~\cite{PRAWER2011}. However, these methods either lack precise spatial control or are non-deterministic, rendering them impractical for data storage. The generation of NV centers by direct laser writing addresses these shortcomings~\cite{EHRLICH1986}. Recent reports have shown the precise positioning of NV centers on the surface~\cite{ZENG2013} and in the bulk~\cite{ROMANO2016,SMITH2017,SMITH2019} by irradiation using NIR pulsed lasers. The charge states of the NV centers (negatively charged NV$^-$ and neutral NV$^0$) can also be controlled by lasers of suitable wavelengths~\cite{WRACHTRUP2013}, using which high-density ODS in diamonds has been demonstrated~\cite{MERILES2016,MERILES2024}. Although ODS using the emission from NV centers can be reversible and highly multiplexed, the instability of the charge states during optical reading has been a limitation for durable data storage under extreme conditions. Recent reports have shown that these limitations can be overcome by using structural defects instead~\cite{DU2024}. In particular, the general radiation1 (GR1) centers are exceptionally stable at high temperatures and are immune to photobleaching. Terabit-scale ODS has been achieved by deterministically laser-writing GR1 centers on the surface and in the bulk of diamond crystals using femtosecond laser pulses~\cite{DU2024}. The stored data are expected to retain their fidelity for millions of years. Although the density of the stored data in diamonds is less than that in dye-doped photoresists~\cite{GU2024-ao}, their durability and lifetime remain unmatched. However, the writing and reading of data using GR1 centers require lasers at two different wavelengths and an elaborate optical setup that includes a confocal microscope. Here, we show that 3D data storage in diamonds using ND1 defects eliminates these drawbacks while maintaining compatibility with the currently used reading and writing systems for ODS.  

\section*{RESULTS}\noindent
\subsection*{Deterministic creation of ND1 defects, writing and reading the data}
\begin{figure} 
	\centering
	\includegraphics[width=0.9\textwidth]{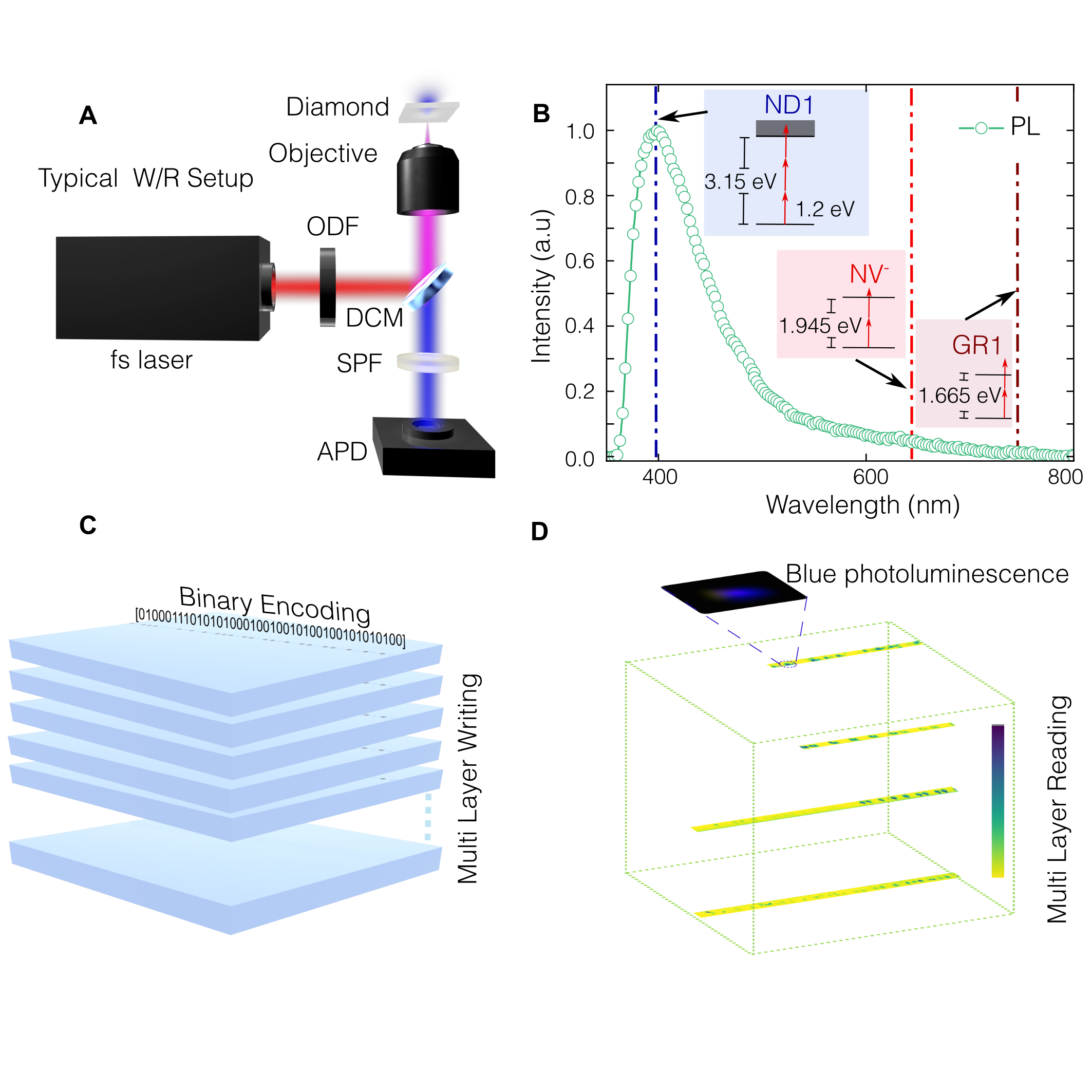} 

	\caption{\textbf{Multilayer optical data writing and readout in diamond.}
		(A) Scheme of a basic read/write device consisting of a femtosecond laser, optical density filter (ODF), a dichoric mirror (DCM), beam-focusing objective, sample on a movable stage, a short-pass filter (SPF) and an avalanche photodiode (APD). (B) Representative PL spectrum of an individual laser-written pit exhibiting a dominant contribution from ND1 centers. Spectral positions of emissive centers that have been used in ODS are indicated by vertical lines. (C) Conceptual illustration of 3D layer-by-layer data encoding in diamond; each dot denotes a single storage unit (pit). (D) Optical readout from four vertically stacked planes at different depths within diamond. Each scan shows distinct emissive pits with intensity represented by the color bar (yellow = no PL, dark magenta = maximum PL). Inset: An optical image of blue PL emitted from a pit. }
	\label{Fig1} 
\end{figure}

\noindent
Sub-micron-sized spots on a diamond crystal with a high density of ND1 defects can be imparted on the surface and in the interior by focusing near-infrared (NIR) femtosecond pulses (for example, femtosecond pulses at 1030 nm) onto the desired location. A simple optical setup consisting of a light source, focusing element, and  stage for accurate positioning of the sample is shown in Fig.~\ref{Fig1}(A). The multiphoton absorption of NIR photons from femtosecond pulses with sufficiently high peak intensities, typically $>$ TW/cm$^2$, and the subsequent avalanche ionization induce structural changes that are localized at the focal point~\cite{JACKMAN2024,ASHIKKALIEVA2022}. ND1 centers, together with many other emissive defects, are formed during this process. The formation of color centers, particularly ND1 centers, is ascertained during the writing process by the appearance of blue emission from the spot. The multiphoton photoluminescence (PL) spectrum depends on the excitation wavelength. When excited with femtosecond pulses at 1030 nm (photon energy 1.2 eV), the spectrum spans the violet to near-infrared (NIR) region. A typical spectrum is shown in Fig. ~\ref{Fig1}(B).
The spectral positions of the emissive centers, in particular NV$^-$ and GR1, which have been used in ODS, are indicated by dashed vertical lines. As the emissions from NV$^-$ and GR1 are at 637 nm (1.945 eV) and 741 nm (1.665 eV), respectively, they can only be induced by multiphoton absorption (MPA) of the laser beam. The energy requirement implies a minimum of two-photon absorption (2PA). The peak below 400 nm originates from the excitation of ND1 centers~\cite{ZAITSEV2022} by the absorption of a minimum of three photons. Remarkably, although the excitation of ND1 centers is by three-photon absorption (3PA), it dominates emissions from other centers, including NV and GR1 centers, indicating a preponderance of ND1 centers compared to other defects. For application in ODS, the peak wavelength in the violet region offers advantages in diffraction-limited resolution while reading the data compared to PL in the visible and near-infrared (NIR) wavelengths~\cite{GU2024-ao,DU2024}. 

The schematic in Fig.~\ref{Fig1}(C) and slices of images in (D) illustrate multilayer data writing and reading. The binary information is encoded as emissive ``pits" on the surface and in the interior of the diamond. The writing is performed using a high-NA objective lens (100x, NA=0.9) at pulse energies in the range of 50-100 nJ. The pits are written sequentially. The location of a pit in the plane is controlled by a nanometer-precision motorized XY stage, whereas its axial position is controlled by moving the objective with a Z-drive. The size of the spot in which the ND1 centers are present can be controlled by the exposure time or energy of the laser pulse. Information retrieval is performed using the same optical setup with a reduced pulse energy ($<$5 nJ). The laser beam is raster-scanned on each recorded plane. As the focal volume sweeps across the layers, the pits produce blue PL, whereas the unmodified regions ("lands") remain dark. Fig.~\ref{Fig1}(D) demonstrates an example of successful information retrieval from four distinct layers. Notably, no interlayer crosstalk is observed, even in raw, unprocessed acquisitions, where discrete PL signals from each layer remain well resolved. While the raw data already enable accurate readouts, further optimization through post-processing can enhance the signal fidelity and system performance. The choice of a laser with a center wavelength of 1030 nm for both writing and reading operations is particularly advantageous for 3D optical data storage because of its low scattering, extended transmission range, and deep penetration depth. Moreover, because the PL from ND1 centers is induced by MPA, the excitation is localized within the focal volume, thereby eliminating the need for confocal microscopy for volumetric imaging and reducing the complexity of the previously reported instrumentation~\cite{GU2024-ao,DU2024}.  

\subsection*{Structural and spectral characterization of the individual storage units}
\begin{figure} 
	\centering
	\includegraphics[width=0.9\textwidth]{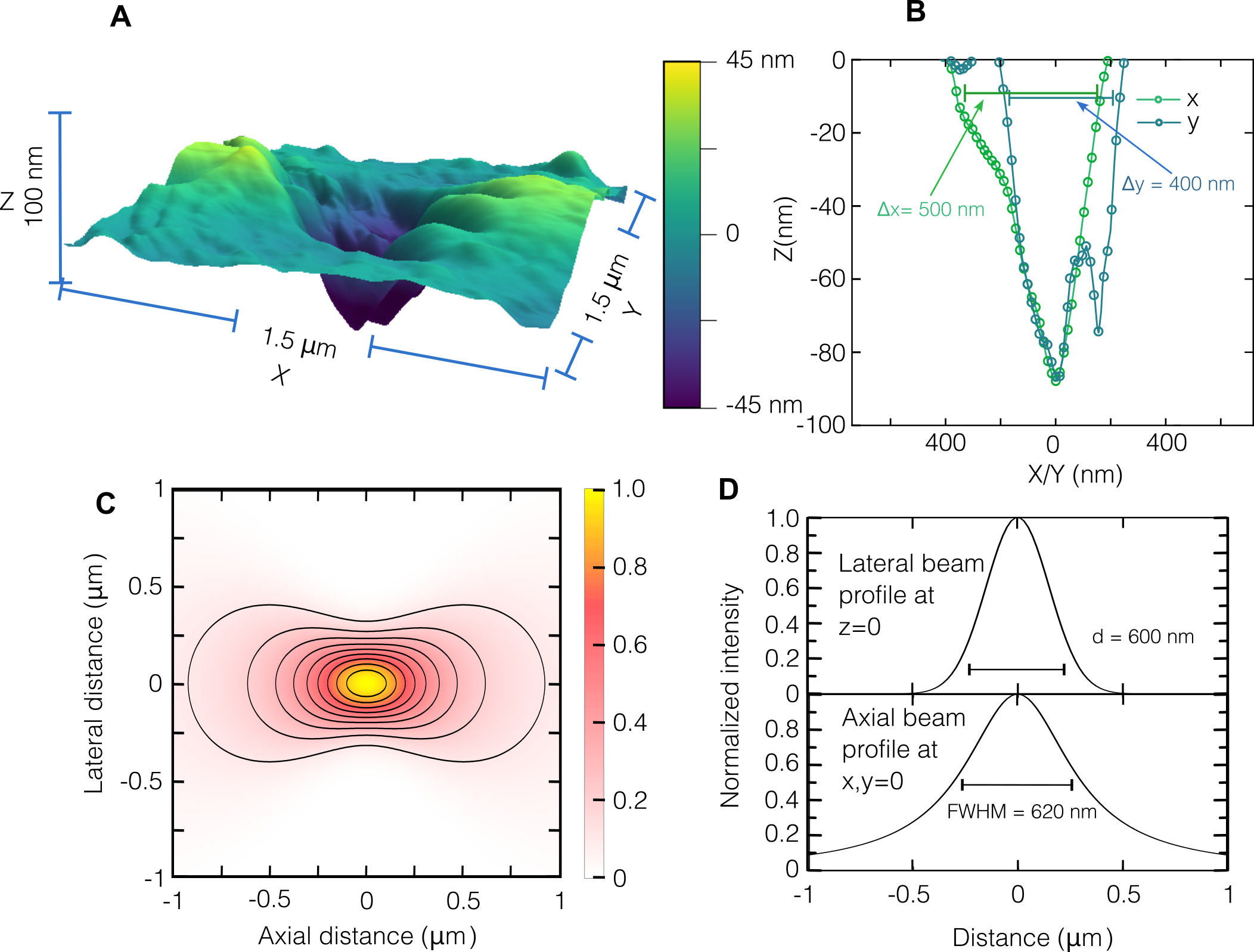} 

	\caption{\textbf{Structural characterization of individual pit.}
(A) AFM image of a single pit, revealing a depth of 90 nm. The pit is not cylindrical. Its width along two perpendicular directions are 400 and 500 nm (B). (C) Simulation of the intensity profile of the beam at the focus spot on the diamond. (D) The beam profile in the lateral and axial directions. The beam has larger widths than the dimensions of the pit. }
	\label{Fig2} 
\end{figure}

\noindent
Atomic force microscopy (AFM) is employed to examine the morphology of a typical pit on the surface in greater detail (Fig.~\ref{Fig2}(A).  The pit is about 90 nm deep. Owing to the highly nonlinear excitation that leads to a large amount of heat deposition during the writing, the lateral shape exhibits some randomness resulting in different widths of 400 and 500 nm in orthogonal directions as shown in Fig.~\ref{Fig2}(B).  Fig.~\ref{Fig2}(C) and (D) show the simulated beam profile for a Gaussian beam with a beam quality of M$^2=1.3$ focused by an objective of NA=0.9. The intensity profile at the focus in the lateral direction is Gaussian with a waist of $w_0 = 300$ nm, and the profile in the axial direction is hyperbolic with a full width at half maximum of 620 nm. Note that the dimensions of the pit are smaller than the focus spot. This is not surprising, considering that multiphoton absorption is necessary to excite electrons in the conduction band, which initiates pit formation. The band gap of diamond (about 5.47 eV) sets the minimum number photons to 5. Thus, pit formation is seeded at the intense part of the focus spot, typically within a radius of $w_0/\sqrt{5}\approx 140$ nm. However, once initiated, the defects absorb photons at a lower intensity, even within a single laser pulse, resulting in the complex growth of pits. It remains to be seen whether pulse shaping and beam pattering can be used to control the dimensions of the pits to lower values.

\begin{figure} 
	\centering
	\includegraphics[width=0.9\textwidth]{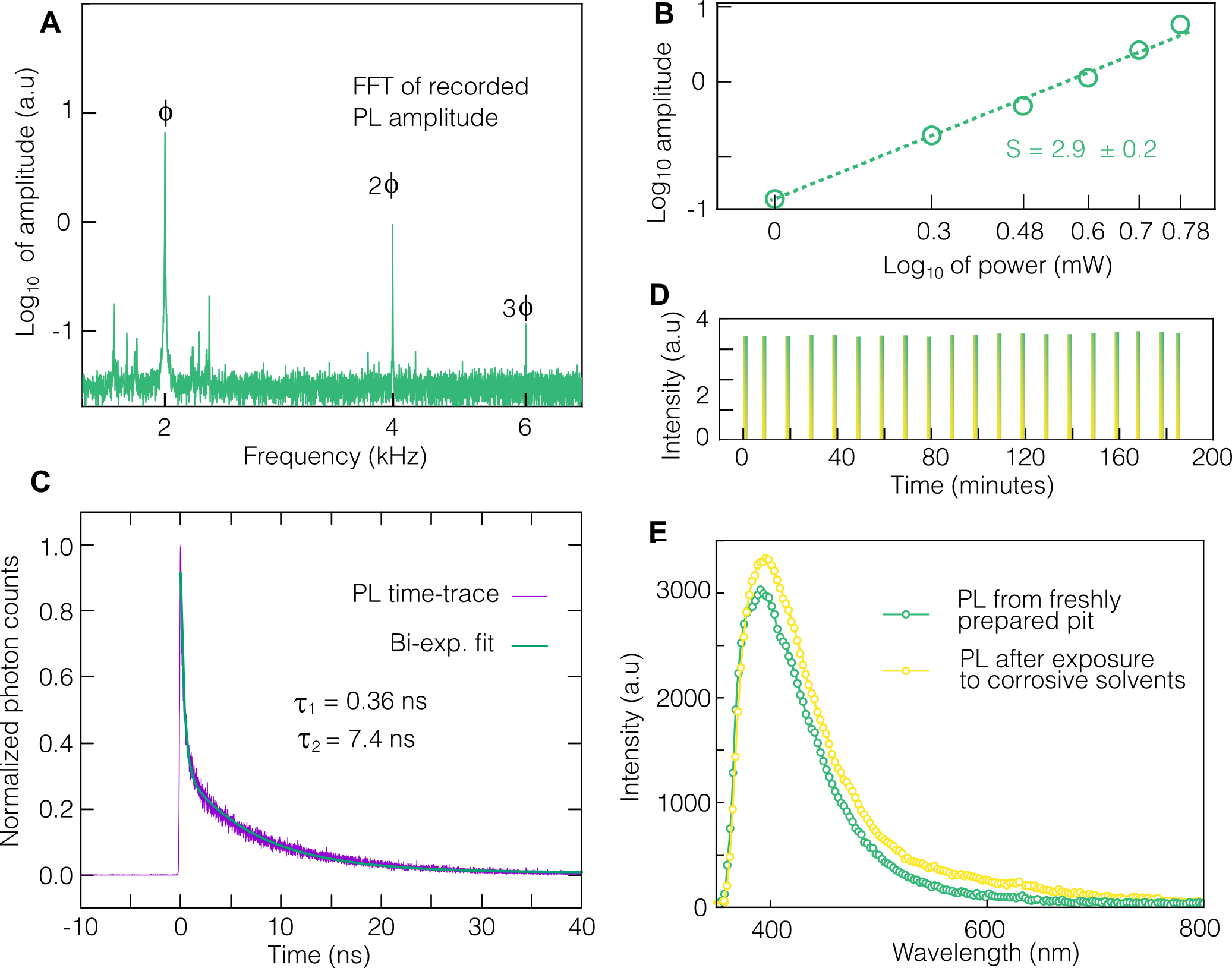} 

	\caption{\textbf{Spectroscopic analysis of PL and excitation process.}
		(A) FFT of the recorded PL amplitude excited by modulated beam at a single frequency $\phi=2$ kHz. The PL shows harmonic distortions at $2\phi$ and $3\phi$ indicating that the excitation is due to 3PA. The cubic dependence of the PL on the average laser power (B) further supports the excitation by 3PA. 
		(C) Time-trace of the PL showing bi-exponential decay. The PL lifetimes are shorter than 10 ns.  (D) The stability of the emission over long exposure to the laser. (E) The spectrum of PL before and after exposure to corrosive environments.}
	\label{Fig3} 
\end{figure}

Excitation by a pair of phase-modulated beams is used to investigate the order of multiphoton absorption that induces PL~\cite{KARKI2016,KARKI2023A,KARKI2024A}. In this method, the carrier frequencies of each beam are shifted slightly and combined collinearly such that the intensity of the resulting beam is modulated at a single frequency ($\phi= 2$ kHz in our measurements). As multiphoton PL is a nonlinear response, it exhibits harmonic distortions based on the order of multiphoton absorption, which can be measured with a high precision. The PL below 420 nm shows additional distortions at $4$ and $6$ kHz (Fig.~\ref{Fig3}(A)), indicating that PL is induced by three-photon absorption (3PA).  The cubic dependence of PL on the excitation intensity as shown in Fig.~\ref{Fig3}(B) provides further support for 3PA. This has several advantages. First, the band-to-band transition in the diamond crystal, which requires 5PA to overcome the bandgap, is not required to excite the ND1 centers, implying that the centers are electronically isolated from the bulk. This is often not the case in other wide-bandgap semiconductors, such as GaN, where luminescent defects are populated by carriers excited to the conduction band~\cite{KARKI2024B}. Second, 3PA requires a lower excitation intensity than 5PA, such that the multiphoton excitation of ND1 centers is feasible without imparting additional defects. Third, at 1030 nm, the combined energy of three photons is 3.6 eV, which is resonant with the ground-to-excited state transitions of the ND1 centers in the spectral region of 3.2-3.6 eV~\cite{ZAITSEV2022} while remaining nonresonant with the transitions in other emissive centers (Fig.~\ref{Fig3}(B)). As can be seen in the PL spectrum (Fig.~\ref{Fig1}(B)), this leads to a dominant contribution from the ND1 centers, while other emissions are minor, simplifying the requirement of spectral filtering of the emission while reading the data. 
Fig.~\ref{Fig3}(C) shows the relaxation kinetics of the emission from the pits measured using time-correlated single-photon counting using the spectrum $<420 $nm. The average lifetime of the PL is $<2 $ ns, which is similar to the lifetime of emission from GR1 centers and molecular systems. The emission lifetime limits the speed at which data can be read when each bits are sequentially scanned in traditional ODS system. With a lifetime of 2 ns, the limit is 500 MBits/s, which can comfortably support the Blu-ray data reading speed of 36 Mbits/s.  

\subsection*{Stability of the emission from the pits}\noindent
First, the photo-stability of the pits is accessed when they are exposed to femtosecond pulses. At normal conditions, no physical change is observed after continuous exposure for 180 min using laser pulses with an energy of $\sim 5$ nJ and duration of $\sim 150$ fs. The PL at the peak ND1 emission wavelength ($\sim 400$~nm) is stable during the measurement as shown in Fig.~\ref{Fig3}(D)). The variation in the intensity of approximately 5\% of the average is higher than the $<2$\% fluctuation of the pulse energy of the laser. This is a consequence of the amplification of fluctuations in nonlinear response rather than other physical processes such as blinking or bleaching. More importantly, the estimated instantaneous 
electric field at the focus spot of $0.3~ \mu$m$^2$ is 
$~1.1\times10^{10}$~ V m$^{-1}$, indicating that 
the stored data withstand extreme transient electric fields. 
Repeated measurements of the PL over a period of a year under normal conditions show no degradation. However, the pit size start to grow if the energy per pulse exceeds $20$ nJ. As discussed previously, although a pulse energy $>50$ nJ is necessary to initiate the formation of defects in our measurements, once formed, they grow at lower pulse energies than this value. Nevertheless, at pulse energies 10 times lower than the threshold for writing, the pits and PL are stable.

Despite the exceptional hardness and chemical inertness of diamonds, color centers in diamonds, such as NV$^0$ and NV$^{-}$, are not stable under extreme physical and harsh chemical conditions~\cite{DU2024}. In this context, if ODS systems based on ND1 centers are to withstand harsh environments, the stability of PL must be investigated after exposure to extreme conditions. The structural and optical stabilities have been accessed under extreme magnetic fields, low and high temperatures, and corrosive chemical environments.   

The chemical durability has been evaluated by immersing the diamond samples in concentrated aqua regia and piranha solutions for 24 h. The post-treatment PL spectra, shown in Fig.~\ref{Fig3}(E), do not show any spectral shift or intensity loss. In fact, a slight increase in the PL is observed, which may be attributed to the cleaning of the surface, resulting in better extraction of PL.  To test the resilience to magnetic fields, samples have been exposed to magnetic fields up to 5 T. In this case also the spectra do not change. Similarly, to test the stability under thermal stress, the samples have been cooled to 4 K and heated to 500 K. Here too, no degradation is observed.

\section*{DISCUSSION}\noindent
In this study, we have demonstrated that ND1 centers can be deterministically created in diamonds at desired locations with sub-micrometer precision using femtosecond pulses. As the characteristic emission in the UV spectral region is stable under extreme electric and magnetic fields, a wide range of temperatures, and highly corrosive environments, they can be used to store data in extreme environments for long periods, extending to millions of years. In the context of ODS, ND1 centers have distinct advantages over other defects. First, the data can be written using Yb fiber-based femtosecond lasers, which are widely used for material processing in industry. This rapidly developing laser technology has shown significant improvements in power, repetition rate, and pulse duration at reduced costs compared with Ti: sapphire lasers. With a demonstrated laser system having a 1.83 GHz repetition rate and pulse energy $>50$ nJ~\cite{LEE2023},  the write speed can reach $>200$ MByte/s, which is higher than the specifications of Blu-ray discs. However, the mechanical movement during writing, which sets the limit for Blu-ray technology, is a bottleneck. Similarly, the read speed is not limited by the laser or by the mechanical movement but by the lifetime of the PL. With a lifetime of 2 ns, a read speed of $~500$ Mbits/s ($\sim 60$ MBytes/s) can be achieved. Second, the wavelength of the Yb-laser is in three-photon resonance with the ground-to-excited state transition of the ND1 centers. In our method, this has been exploited to verify the writing of the data in situ and in real-time. Third, three-photon excitation provides highly localized absorption, using which the reading of the 3D data is accomplished with standard optical elements without requiring confocal microscopy~\cite{DU2024}. Coincidentally, excitation at 1030 nm is not multiphoton-resonant with the common emissive centers in diamonds, including the NV centers and  GR1.  This reduces the contributions from other defects and avoids damage to the pits during data readout. PL from other defects overwhelms when lasers at visible wavelengths are used. 

The use of amplified femtosecond lasers has obvious disadvantages too. With the currently available laser technologies, the system is bulky and energy inefficient, severely limiting its use. Nevertheless, the encouraging advancement of on-chip femtosecond laser amplification is likely to mitigate these disadvantages~\cite{GORODETSKY2018,MARANDI2023,HERR2024}. We have used undoped pure diamond in our study. A study on the effects of doping on the creation of ND1 centers could help to optimize, and ideally lower the pulse energies for reading and writing. 
Similarly, spatio-temporal pulse shaping may yield a smaller pit size comparable to that of Blu-ray technology, resulting in 3D-ODS with data storage capacity rivaling the solid-state drives.  It should be noted that the pits cannot be erased under normal conditions; thus, the medium is not rewritable. In this context, the system is ideally suitable for archiving important data over extended periods, even when exposed to extreme conditions. 

\section*{EXPERIMENTAL SETUPS}\noindent
\subsection*{Setup for writing/reading}\noindent
Single-crystal diamonds from Element Six with nitrogen impurity concentrations below 5 ppb and NV concentrations below 0.03 ppb were used as ODS samples. A schematic of the read/write setup is shown in Fig.~\ref{Fig4}(A). A femtosecond pulsed laser ( Impulse from Clark-MXR, Inc. ) operated at a repetition rate of 1 MHz, central wavelength of 1030 nm (photon energy ~1.2 eV), and pulse duration of ~150 fs served as the source for both the writing and reading operations. The beam was directed to a high-numerical-aperture objective lens (Nikon, 100×, NA = 0.9) using an 800 nm short-pass dichroic mirror (Edmund Optics, \#69-220). A high-speed shutter was used to control the laser exposure. The beam, which exhibited a near-perfect Gaussian profile, was tightly focused into the diamond medium through the objective. The incident laser intensity was regulated using a continuously variable neutral-density filter. The diamond sample was mounted on a high-precision motorized X-Y nanopositioning stage (HLD117NN, Prior Scientific). The axial position of the focus spot was controlled by moving the objective with an independent motorized Z-axis drive (PS3H122R, Prior Scientific). The PL was collected in the epi-direction by the same objective. It was directed to an avalanche photodiode (APD) after passing through an additional 800 nm short-pass filter to suppress the residual excitation light. The signal from the APD was digitized using a 24-bit acquisition card. All optical signals and motorized scanning stages were synchronized using custom-developed software. 

\subsection*{Measurement of PL liftime}\noindent
The same optical setup for reading/writing was used for the measurements of the PL lifetime. In place of the APD, a single-photon counting avalanche photodiode (SPAD) was used. A time-tagger (Time Tagger 20, Swabian instruments) was used to record the photon counts and histogram them.  

\subsection*{Setup for intensity modulation}
\begin{figure} 
	\centering
	\includegraphics[width=0.9\textwidth]{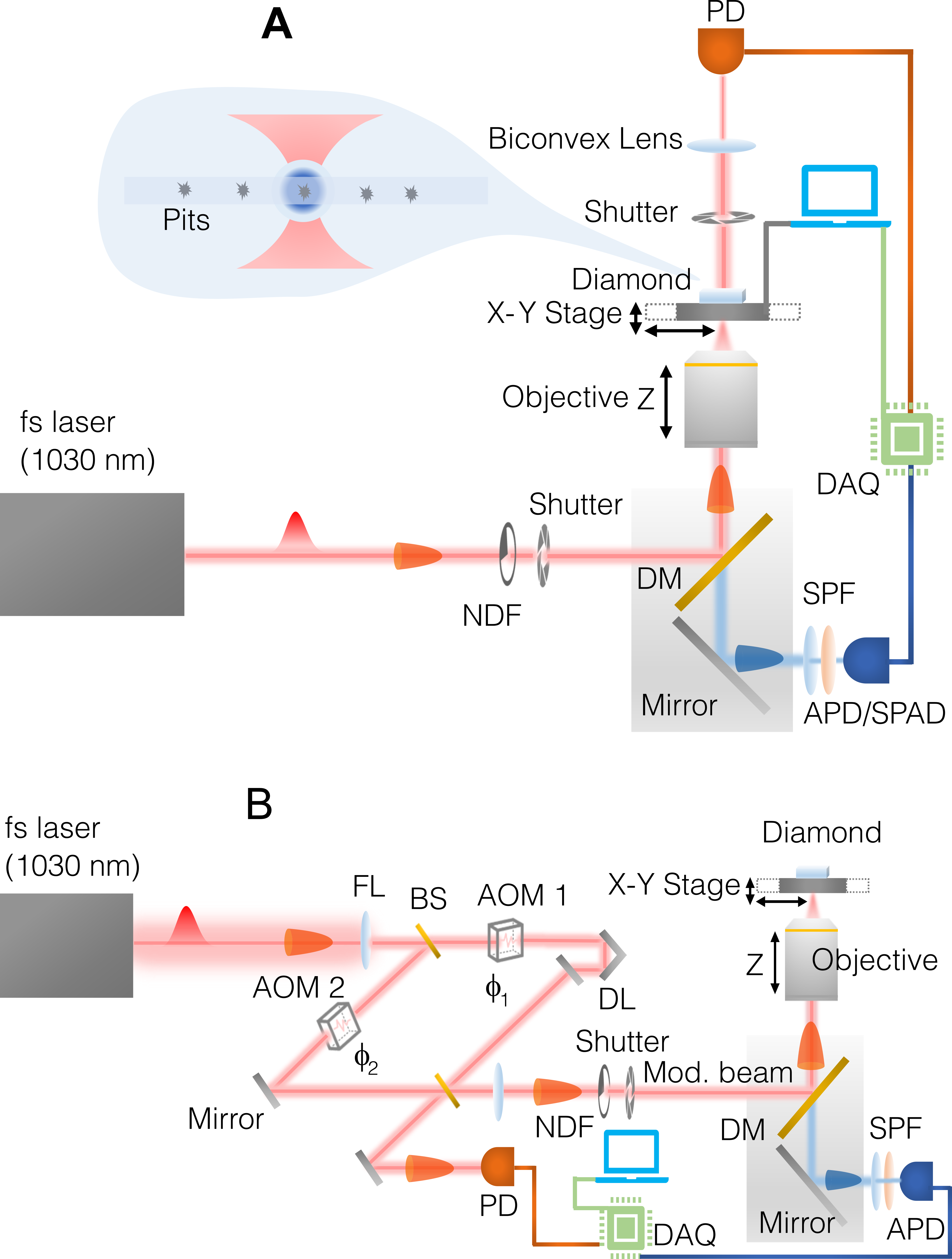} 

	\caption{\textbf{Detailed experimental and characterization setups.}
		(A) Schematic of the optical setup used for data writing and reading in diamond. NDF: Neutral density filter; DM: Dichoric mirror (short pass); PD: Photodiode; SPF: Short-pass filter; APD: Avalanche photodiode; SPAD: Single photon counting avalanche photodiode; DAQ: Data acquisition card. (B) Setup for phase modulation employed to characterize the nonlinear excitation processes responsible for PL. FL: Focusing lens; BS: 50/50 Beam splitter; AOM: Acousto-optic modulator; DL: Piezo-driven delay line.}
	\label{Fig4} 
\end{figure}

\noindent
A home-built Mach–Zehnder interferometer, schematically illustrated in Fig.~\ref{Fig4}(B), was employed to obtain a clean modulation of the average power of the laser beam at a single frequency. The output of the femtosecond laser was equally split into two arms using a non-polarizing beam splitter. Each arm was independently directed to an acousto-optic modulator (AOM). The AOMs were driven at two different frequencies, $\phi_1$ and $\phi_2$, referenced to a single clock. The carrier frequencies of the diffracted outputs from the AOMs were shifted to $\omega_1=\omega+\phi_1$ and $\omega_2=\omega+\phi_2$, where $\omega$ is the carrier frequency of the laser. The two frequency-shifted beams were subsequently overlapped collinearly in space and time, producing a clean intensity modulated beam at the frequency $\phi=\phi_2-\phi_1=2 $ kHz. The intensity-modulated beam was tightly focused on the diamond sample to induce blue PL. A silicon photodetector was placed at second output of the interferometer to monitor the modulated beam as the reference. The PL was collected in the same way as when reading the data. Both signals were digitized using a 24-bit acquisition card for synchronized detection and analyses. 

	\centering
	


\clearpage 

%
\bibliography{science_template} 
\bibliographystyle{sciencemag}

%
%
%
%
%
%


\section*{Acknowledgments}
The authors thank Prof. Elissaios Stavrou of Guangdong Technion-Israel Institute of Technology for providing a diamond crystal as a sample.

\paragraph*{Funding:}
KJK acknowledges financial support from the National Key Research and Development Program of China (grant no. 2023YFA1407100), Guangdong Province Science and Technology Major Project (Future functional materials under extreme conditions - 2021B0301030005), and Li Ka Shing Foundation STU-GTIIT Joint Research Grants (2024LKSFG02). 
\paragraph*{Author contributions:}
KJK conceived the study. AA and WH performed the measurements and data analysis. AA and KJK drafted the manuscript.

\newpage

\end{document}